\documentclass[twocolumn,prb,amsmath,amssymb,floatfix]{revtex4}
\pdfoutput=1
\usepackage{graphicx}
\usepackage{dcolumn}
\usepackage{bm}

\begin{document}

\title{Single-shot readout of an electron spin in silicon}

\author{Andrea Morello$^{1*}$, Jarryd J. Pla$^{1}$, Floris A. Zwanenburg$^{1}$, Kok W. Chan$^{1}$, Hans Huebl$^{1\dag}$, Mikko M\"{o}tt\"{o}nen$^{1,3,4}$, Christopher D. Nugroho$^{1\ddag}$, Changyi Yang$^{2}$, Jessica A. van Donkelaar$^{2}$, Andrew D. C. Alves$^{2}$, David N. Jamieson$^{2}$, Christopher C. Escott$^{1}$, Lloyd C. L. Hollenberg$^{2}$, Robert G. Clark$^{1}$, and Andrew S. Dzurak$^{1}$}

\affiliation{
$^1$ Centre for Quantum Computer Technology, School of Electrical Engineering \& Telecommunications, {University of New South Wales, Sydney NSW 2052, Australia.}\\
$^2$ Centre for Quantum Computer Technology, School of Physics,
University of Melbourne, Melbourne VIC 3010,
Australia.\\
$^{3}$ Department of Applied Physics/COMP, Aalto University, P.O.
Box 15100, 00078 Aalto, Finland.\\
$^{4}$ Low Temperature Laboratory, Aalto University, P.O. Box 13500,
00078 Aalto, Finland.\\}

\maketitle

\textbf{The size of silicon transistors used in microelectronic
devices is shrinking to the level where quantum effects become
important \cite{levi07IEEE}. While this presents a significant
challenge for the further scaling of microprocessors, it provides
the potential for radical innovations in the form of spin-based
quantum computers \cite{loss98PRA,kane98N,hollenberg06PRB}  and
spintronic devices \cite{zutic04RMP}. An electron spin in Si can
represent a well-isolated quantum bit with long coherence times
\cite{tyryshkin03PRB} because of the weak spin-orbit coupling
\cite{feher59PR} and the possibility to eliminate nuclear spins from
the bulk crystal \cite{ager05JEC}. However, the control of single
electrons in Si has proved challenging, and has so far hindered the
observation and manipulation of a single spin. Here we report the
first demonstration of single-shot, time-resolved readout of an
electron spin in Si. This has been performed in a device consisting
of implanted phosphorus donors \cite{jamieson05APL} coupled to a
metal-oxide-semiconductor single-electron transistor
\cite{angus07NL, morello09PRB} -- compatible with current
microelectronic technology. We observed a spin lifetime approaching
1 second at magnetic fields below 2 T, and achieved spin readout
fidelity better than 90\%. High-fidelity single-shot spin readout in
Si opens the path to the development of a new generation of quantum
computing and spintronic devices, built using the most important
material in the semiconductor industry.}

The projective, single-shot readout of a qubit is a crucial step in
both circuit-based and measurement-based quantum computers
\cite{ladd10N}. For electron spins in solid state, this has only
been achieved in GaAs/AlGaAs quantum dots coupled to charge
detectors \cite{elzerman04N,hanson05PRL, barthel09PRL}. The spin
readout was achieved utilizing spin-dependent tunnelling, in which
the electron was displaced to a different location depending on its
spin state. The charge detector, electrostatically coupled to the
electron site, sensed whether the charge had been displaced, thereby
determining the spin state. Here we apply a novel approach to charge
sensing, where the detector is not only electrostatically coupled,
but also tunnel-coupled to the electron site \cite{morello09PRB}, as
shown in Fig.~1a. As a charge detector we employ here the silicon
single-electron transistor\cite{angus07NL} (SET), a nonlinear
nanoelectronic device consisting of a small island of electrons
tunnel-coupled to source and drain reservoirs, electrostatically
induced beneath an insulating SiO$_2$ layer. A current can flow from
source to drain only when the electrochemical potential of the
island assumes specific values\cite{devoret00N}, resulting in a
characteristic pattern of sharp current peaks as a function of gate
voltage (Fig.~1e). The shift in electrochemical potential arising
from the tunnelling of a single electron from a nearby charge centre
into the SET island is large enough to switch the current from zero
to its maximum value. This tunnelling event becomes spin-dependent
in the presence of a large magnetic field, when the spin-up state
$|\uparrow \rangle$ has a higher energy than the spin-down state
$|\downarrow \rangle$, by an amount larger than the thermal and
electromagnetic broadening of electron states in the SET island.
Therefore we perform the experiment in high magnetic fields, $B >
1$~T, and very low electron temperatures, $T_{\rm el} \approx
200$~mK.

The high effective mass and the valley degeneracy in silicon
\cite{goswami07NP} require very tight confinement to isolate a
single electron in a non-degenerate state. Phosphorus atoms in
silicon naturally provide a sharp confining potential for their
bound donor electron, with the additional advantage that the
$^{31}$P nuclear spin can be used as a long-lived quantum memory
\cite{morton08N}. Therefore we have fabricated a device where P
donors were implanted in a small region ($90 \times 90$~nm$^2$) next
to the SET (Fig.~1c). The P$^+$ ion fluence was chosen to maximize
the likelihood that 3 donors are located at a distance $\sim 30 -
60$~nm from the SET island and can be tunnel-coupled to it, forming
a parallel double-quantum-dot system \cite{hofmann95PRB}. The SET
top gate and a plunger gate overlaying the P implanted area provide
full electrostatic control of the hybrid double-dot.

\begin{figure}[b] \center
\includegraphics[width=7.2cm]{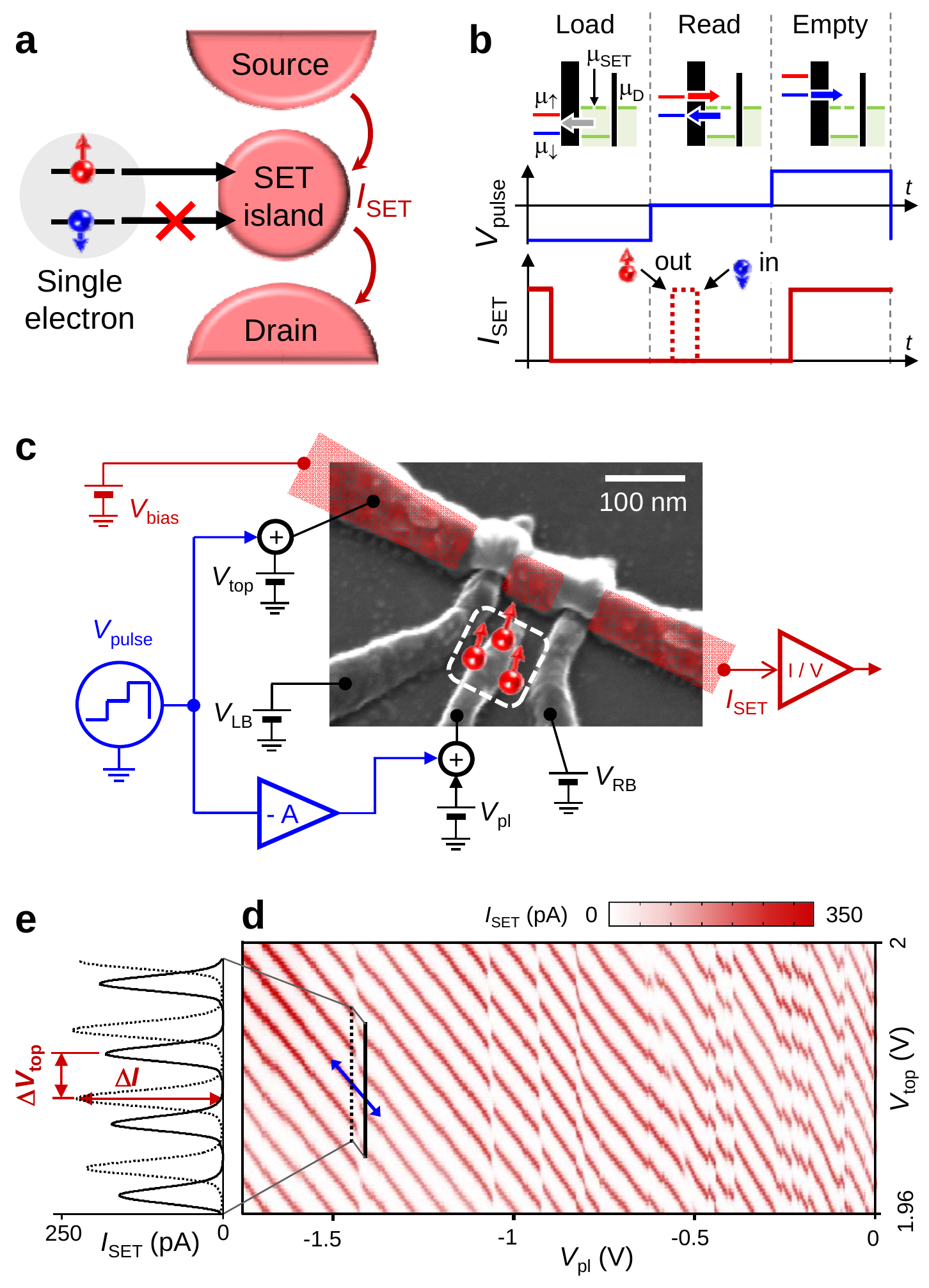}
\caption{ \label{Fig1} \textsf{\textbf{Spin readout device
configuration and charge transitions. a} Schematic showing the
spin-dependent tunnelling configuration, where a single electron can
tunnel onto the island of a SET only when in a spin-up state.
\textbf{b} Pulsing sequence for single-shot spin readout (see main
text), and SET response, $I_{\rm SET}$. The dashed peak in $I_{\rm
SET}$ is the expected signal from a spin-up electron. The schematics
at the top depict the electrochemical potentials of the electron
site ($\mu_{\downarrow,\uparrow}$), of the SET island ($\mu_{\rm
SET}$) and of the drain contact ($\mu_{\rm D}$). \textbf{c} Scanning
electron micrograph of a device similar to the one measured. The P
donors implant area is marked by the dashed square. The total
expected number of implanted donors is 18 (Poisson statistics), of
which $\sim 3$ are close enough to the SET to be significantly
tunnel-coupled. Both DC voltages and pulses are applied to the gates
as indicated. The red shaded area represents the electron layer
induced by the top gate and confined beneath the SiO$_2$ gate oxide
layer. \textbf{d} SET current $I_{\rm SET}$ as a function of the
voltages on the top and the plunger gates, $V_{\rm top}$ and $V_{\rm
pl}$, at $B = 0$. The lines of SET Coulomb peaks are broken by
charge transfer events. The blue arrow on the transition at $V_{\rm
pl} \approx -1.4$~V shows the axis along which $V_{\rm top}$ and
$V_{\rm pl}$ are pulsed for compensated time-resolved measurements,
ensuring that $\mu_{\rm SET}$ remains constant during the pulsing.
\textbf{e} Line traces of $I_{\rm SET}$ along the solid and dashed
lines in panel \textbf{d}. Ionizing the donor shifts the sequence of
SET current peaks by an amount $\Delta V_{\rm top} = \Delta q/C_{\rm
top}$, causing a change $\Delta I$ in the current. The charging
energy of the SET is $E_{\rm C} \sim 1.5$~meV.}}
\end{figure}

To perform spin readout we bias the gates so as to tune the
electrochemical potentials on the SET ($\mu_{\rm SET}$) and a nearby
donor ($\mu_{\downarrow}$ and $\mu_{\uparrow}$ for states
$|\downarrow \rangle$ and $|\uparrow \rangle$, respectively) such
that the SET current, $I_{\rm SET}$, is zero when the electron
resides on the donor, while $I_{\rm SET}\neq 0$ when the donor is
ionized. The readout protocol consists of three
phases\cite{elzerman04N}, shown in Fig.~1b. (i) A Load phase, during
which an electron in an unknown spin state tunnels from the SET
island to the donor, since $\mu_{\rm SET} >
\mu_{\downarrow},\mu_{\uparrow}$. The electron loading is signalled
by $I_{\rm SET}$ dropping to zero. (ii) A Read phase, during which a
spin-down electron remains trapped on the donor, leaving $I_{\rm
SET} = 0$, but a spin-up electron can tunnel onto the SET island,
causing $I_{\rm SET} = I_{\rm max}$. A (different) spin-down
electron from the SET island can later tunnel back onto the donor,
blocking the current again. Therefore, the signal of a state
$|\uparrow \rangle$ is a single current pulse at the beginning of
the Read phase. (iii) An Empty phase during which the donor is
ionized, to ensure a new electron with random orientation can be
loaded at the next cycle.

Measuring $I_{\rm SET}$ as a function of the plunger ($V_{\rm pl}$)
and SET top ($V_{\rm top}$) gate voltages, yields the map shown in
Fig.~1d. Each time a charge centre coupled to the SET changes its
charge state, the sequence of SET current peaks breaks and shifts in
gate voltage by an amount $\Delta V_{\rm top} = \Delta q/C_{\rm
top}$, where $\Delta q$ is the charge induced on the SET island and
$C_{\rm top}$ is the capacitance between island and top gate
(Fig.~1e). Figure 1d shows a large number of charge transitions for
$-0.7 < V_{\rm pl} < 0$~V. Most of these transitions are
irreproducible and hysteretic, and are probably caused by the
charging/discharging of shallow traps at the Si/SiO$_2$ interface.
The transitions for $V_{\rm pl} < -0.7$~V, however, are stable and
well reproduced even after several thermal cycles. Their number is
consistent with the expected number of implanted donors in the
active area, and they have been observed in a number of similar
devices \cite{huebl09CM}. Considering the results of the spin
lifetime measurements discussed below, it is likely that we are
observing transitions between $D^+$ and $D^0$
states\cite{sellier06PRL} of implanted P donors\cite{tan10NL}.

\begin{figure}[b] \center
\includegraphics[width=8.7cm]{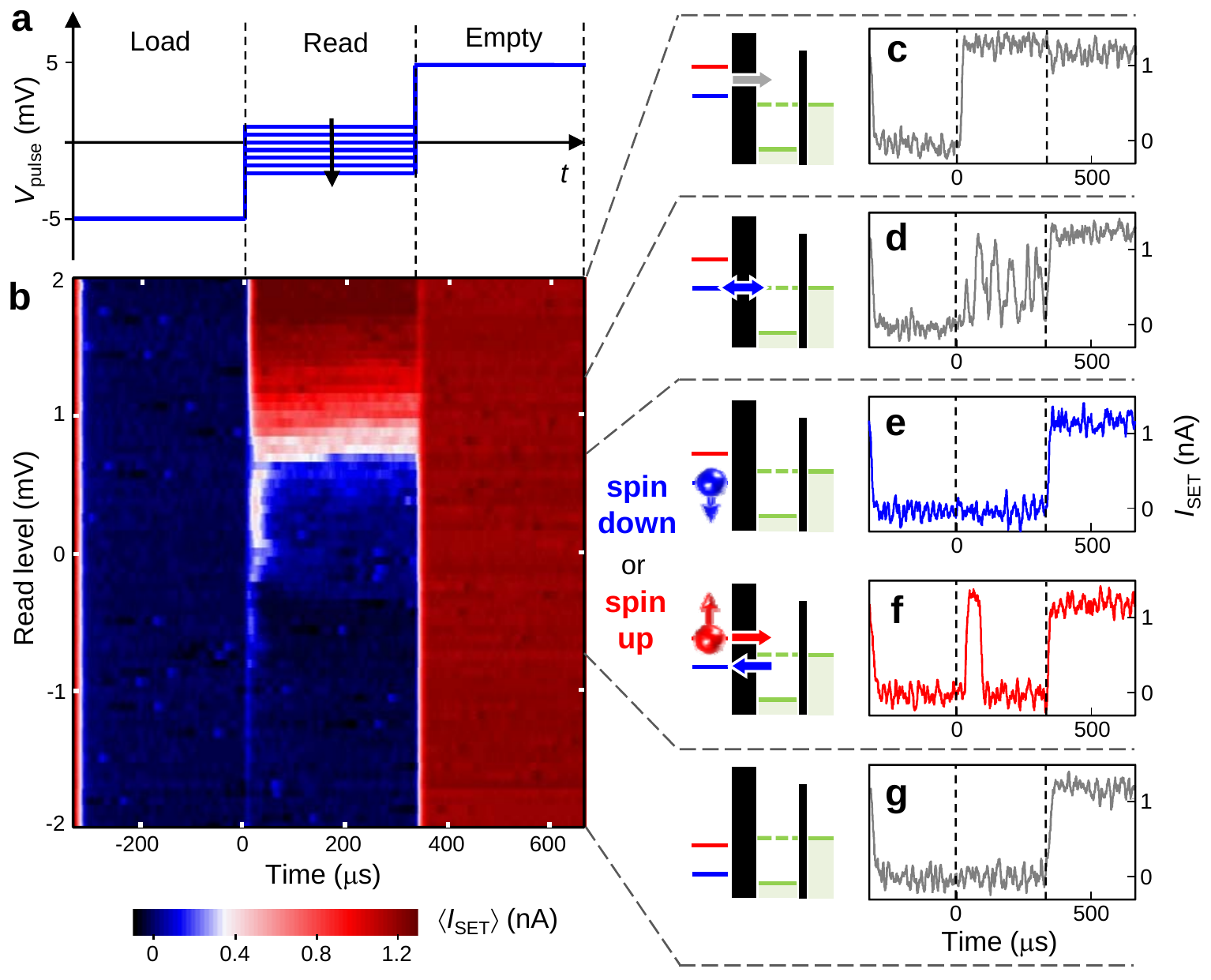}
\caption{ \label{Fig2} \textsf{\textbf{Single-shot spin readout and
calibration of the read level. a} Three-level pulsing sequence for
spin readout. The Load and Empty levels are kept constant, while the
Read level is scanned from high to low. \textbf{b} SET current
$I_{\rm SET}$ , averaged over 128 single-shot traces, as a function
of the $V_{\rm pulse}$ level during the Read phase. Data taken with
an applied magnetic field $B = 5$~T and a detection bandwidth of 40
kHz (rise time $~10$~s). \textbf{c - g} Examples of single-shot
traces. \textbf{c} Read level too high, $\mu_{\downarrow,\uparrow}
> \mu_{\rm SET}$: The electron always leaves the donor during the Read pulse,
regardless of its spin. \textbf{d} $\mu_{\downarrow} \approx
\mu_{\rm SET}$: Random telegraph signal indicates an electron
switching between SET island and $|\downarrow \rangle$ state.
\textbf{e and f} Correct Read level, $\mu_{\downarrow} < \mu_{\rm
SET} < \mu_{\uparrow}$: $I_{\rm SET}$ = 0 during the Read phase
indicates a $|\downarrow \rangle$ state (e). A single current pulse
at the beginning of the Read phase is the signature of a $|\uparrow
\rangle$ electron (f). The regime of correct Read level is
recognizable by the isolated increase in $I_{\rm SET}$. \textbf{g}
Read level too low, $\mu_{\downarrow,\uparrow} < \mu_{\rm SET}$: The
electron never leaves the donor during the Read pulse.}}
\end{figure}

The charge transition at $V_{\rm pl} \approx -1.4$~V in Fig.~1d has
a large $\Delta q \approx 0.7e$, where $1e$ is equivalent to the
spacing between adjacent current peaks. This indicates a donor very
close to the SET island \cite{morello09PRB}. Accordingly, we find a
fast electron tunnelling time between the donor and the SET, of
order 10~$\mu$s. For comparison, the charge transition at $V_{\rm
pl} \approx -1.1$~V has a lower $\Delta q \approx 0.3e$ and a much
slower tunnel time $\sim 10$~ms, consistent with a donor further
away. We chose the donor transition at $V_{\rm pl} \approx -1.4$~V
to implement the spin readout protocol. Figure 2b-g illustrates the
method we used to find the values of $V_{\rm pulse}$ during the Read
phase at which spin-dependent tunnelling is achieved. By lowering
the Read level from too high (Fig.~2c) to too low (Fig.~2g), the
time traces of $I_{\rm SET}$ during the Read phase show a transition
from $I_{\rm SET} = I_{\rm max}$, through random telegraph signal,
to $I_{\rm SET}= 0$, passing through a region where $I_{\rm SET}$
can be either zero (Fig.~2e) or show a spin-up signal (Fig.~2f). In
this region, the condition $\mu_{\downarrow} < \mu_{\rm SET} <
\mu_{\uparrow}$ is fulfilled, and a single-shot projective
measurement of the electron spin state is performed. When plotting
the average of several single-shot traces taken at different read
levels, the correct readout range is highlighted by the appearance
of a high current region at the beginning of the Read phase,
spanning a time interval of the order of the electron tunnel time
(see also Fig.~3c). Such a high-current region is absent in
measurements performed in zero magnetic field, as expected. With a
modified pulse sequence, it is also possible to extract the Zeeman
energy splitting $E_{\rm Z} = g\mu_{\rm B}B$ and demonstrate the
deterministic loading of a $|\downarrow \rangle$ electron (see
Supplementary information). Because the loading of a state
$|\downarrow \rangle$ is controlled by gate voltages and occurs on
$\sim 10$~$\mu$s time scales as determined by the electron tunnel
time, this device already realizes two essential requirements for
quantum computation and quantum error correction, namely, the
single-shot readout and the fast preparation of the qubit ground
state\cite{divincenzo00FP}.

For each single-shot measurement, the state is identified as
$|\uparrow \rangle$ when $I_{\rm SET}$ surpasses a suitably chosen
threshold $I_{\rm T}$ (Fig.~3b) during the first 100 $\mu$s of the
Read phase. Defining $P_{\uparrow}$ as the probability to observe a
spin-up electron, we find that $P_{\uparrow}$ decreases when
increasing the wait time $\tau_{\rm w}$ before the spin is read out
(Fig.~3a), because the  excited state $|\uparrow \rangle$ relaxes to
the ground state $|\downarrow \rangle$. The wait time dependence of
$P_{\uparrow}$  (Fig.~3d) is well described by a single exponential
decay, $P_{\uparrow} (\tau_{\rm w}) = P_{\uparrow}(0)\exp(-\tau_{\rm
w}/T_1)$, where $T_1$ is the excited state lifetime. We note that
measuring $T_1$ does not strictly require single-shot
readout\cite{hayes09CM,xiao10PRL}. Since the spin-up current pulses
occur with the highest probability in a well defined time interval,
the average current $\langle I_{\rm SET} \rangle(t)$ has a
Poissonian shape (Fig.~3c), and its integral is proportional to
$P_{\uparrow}$. Figure 3d shows that the integral of $\langle I_{\rm
SET} \rangle(t)$ for $0 < t < 100$~$\mu$s can be rescaled and
superimposed with $P_{\uparrow}$ as obtained from single-shot
readout, and an exponential fit yields the same $T_1$ for both
methods. Single-shot readout provides an absolute measure of
$P_{\uparrow}$., but integrating $\langle I_{\rm SET} \rangle(t)$ is
simpler and faster, and is used for the $T_1$ measurements below.

\begin{figure}[t] \center
\includegraphics[width=7.8cm]{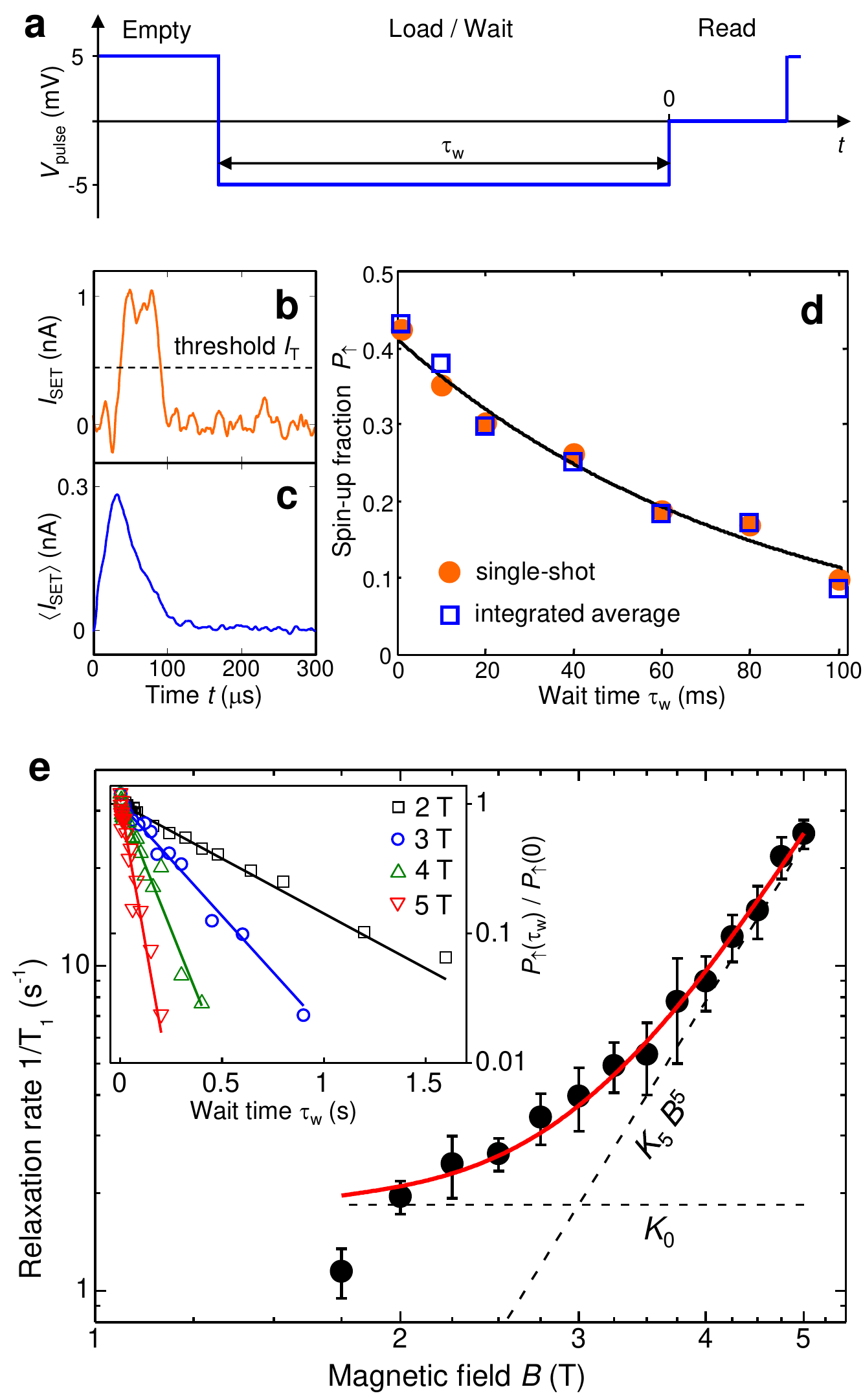}
\caption{ \label{Fig3} \textsf{\textbf{Spin-lattice relaxation rate.
a} Pulsing sequence for measuring the spin relaxation rate $1/T_1$,
identical to Fig.~2a but with a variable Load/Wait time $\tau_w$.
\textbf{b} Single-shot spin readout traces. A $|\uparrow \rangle$
state is counted when $I_{\rm SET} > I_{\rm T} = 0.45$~nA.
\textbf{c} Average of 500 single-shot traces. \textbf{d} The spin-up
fraction $P_{\uparrow}(\tau_{w})$ can be obtained by counting the
$|\uparrow \rangle$ states in single-shot (dots) or by integrating
$\langle I_{\rm SET} \rangle(t)$ (squares). Rescaling and
normalizing the integral of $\langle I_{\rm SET}\rangle$ shows that
the two approaches are equivalent and can be fitted by the same
exponential decay (solid line). Here $B = 5$~T. \textbf{e} Magnetic
field dependence of $1/T_1$. Error bars are 95\% confidence levels.
The data follow $1/T_1(B) = 1.84$~s$^{-1}$~+~$0.0076
B^5$~s$^{-1}$T$^{-5}$ (red line, sum of the dashed lines). The point
at $B = 1.75$~T is not included in the fitted dataset. Inset:
exponential decays of the normalized spin-up fraction, at different
magnetic fields as indicated.}}
\end{figure}

The measured spin relaxation rate as a function of magnetic field,
$T_1^{-1}(B)$, at phonon temperature $T \approx 40$~mK, is plotted
in Fig.~3e. The data for $B \geq 2$~T are well described by the
function $T_1^{-1}(B)\approx K_0 + K_5 B^5$, with $K_0 = 1.84 \pm
0.07$~s$^{-1}$ and $K_5 = 0.0076 \pm 0.0002$~s$^{-1}$T$^{-5}$.   A
fit of the form $T_1^{-1}(B) = K_0 + K_a B^a$, where $K_0$, $K_a$
and $a$ are free parameters, yields $a = 4.8 \pm 0.2$. The $B^5$
dependence agrees with the low-$T$ limit\cite{hanson07RMP} ($k_{\rm
B}T \ll g\mu_{\rm B}B$) of a spin-lattice relaxation mechanism
arising from the effect of spin-orbit coupling, through the
deformation of the crystal lattice caused by the emission of an
acoustic phonon\cite{hasegawa60PR} when the state $|\uparrow
\rangle$ relaxes to $|\downarrow \rangle$. This result is
incompatible with the known relaxation process for interface traps,
which is dominated by the coupling to two-level
fluctuators\cite{desousa07PRB}. A recent electron spin resonance
experiment on shallow traps at the Si/SiO$_2$
interface\cite{shankar09CM} found $T_1 \sim 800$~$\mu$s at $T =
350$~mK and $B = 0.32$~T, i.e. 2 to 3 orders of magnitude shorter
than our result, despite the much lower magnetic field. Conversely,
our measurements are in qualitative and quantitative agreement with
the behaviour of $T_1^{-1}$ measured in bulk-doped P:Si
samples\cite{feher59PR} (see Supplementary information for a
detailed discussion). This constitutes a strong indication that we
have measured the spin of a single electron bound to an implanted P
donor.

\begin{figure}[t] \center
\includegraphics[width=8.7cm]{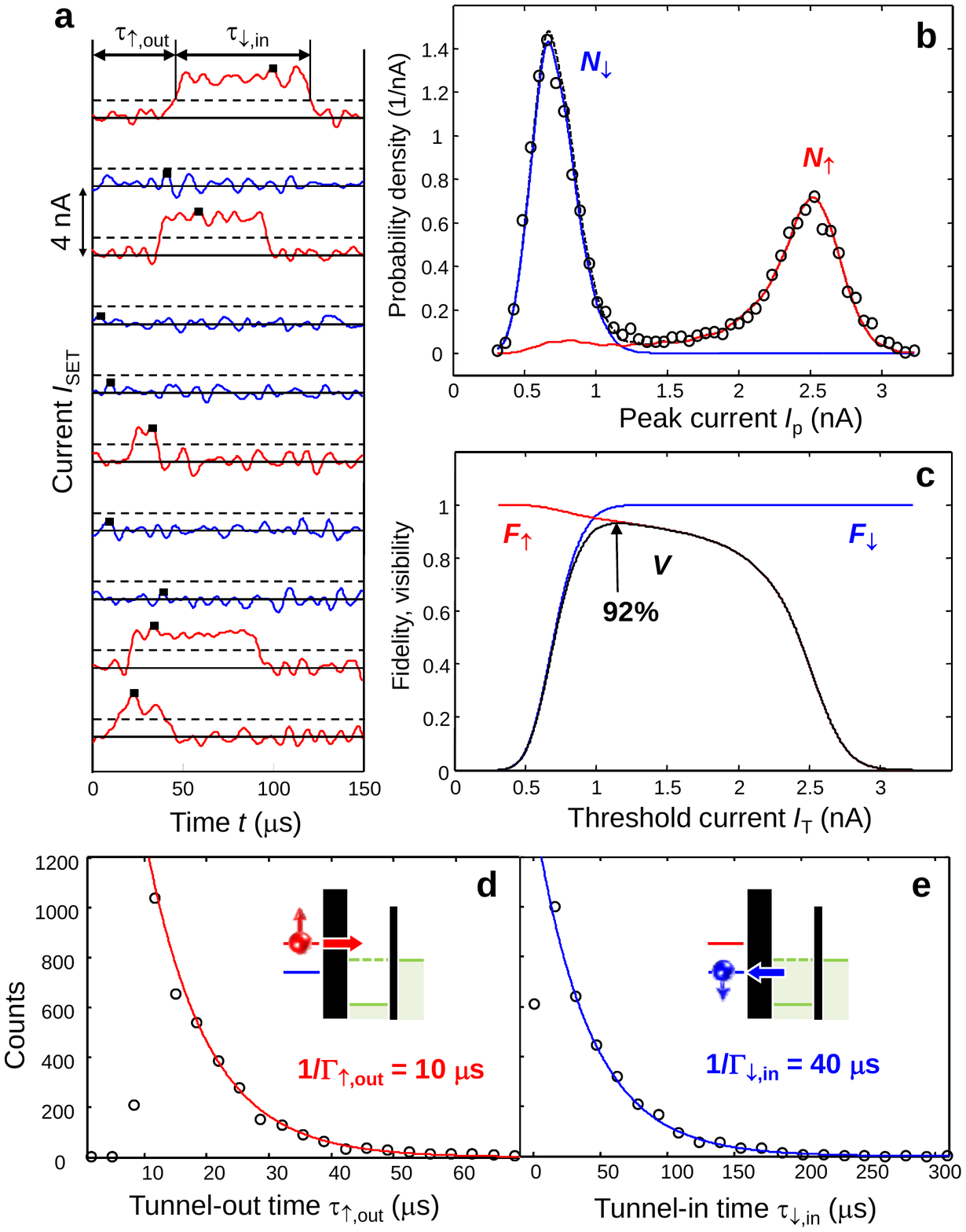}
\caption{ \label{Fig4} \textsf{\textbf{Readout fidelity and
visibility. a} Examples of single-shot $I_{\rm SET}$ traces, each
shifted by 4 nA for clarity, with $B = 5$~T and 120 kHz bandwidth
($\sim 3$~$\mu$s rise/fall time). The spin is labeled $|\uparrow
\rangle$ (red) or $|\downarrow \rangle$ (blue) depending on whether
$I_{\rm SET}$ passes the threshold $I_{\rm T} = 1.1$~nA (dashed
lines). \textbf{b} Histogram (circles) of the maximum values of
$I_{\rm SET}$ in the interval $0 < t < 100$~$\mu$s (black squares in
panel \textbf{a}), obtained from a 10,000 shots dataset. The blue
and red lines are simulated histograms for states $|\downarrow
\rangle$ and $|\uparrow \rangle$, respectively, and the black dashed
line is the sum of the two. The simulated curves are obtained using
$P_{\uparrow} = 0.47$, $\Delta I = 1.9$~nA,
$1/\Gamma_{\uparrow,\mathrm{out}} = 10$~$\mu$s,
$1/\Gamma_{\downarrow,\mathrm{in}} = 40$~$\mu$s. \textbf{c}
$|\downarrow \rangle$ (blue) and $|\uparrow \rangle$ (red) readout
fidelities, and readout visibility (black) as a function of the
discrimination threshold $I_{\rm T}$. The maximum visibility is 92\%
at $I_{\rm T} \approx 1.1$~nA. \textbf{d and e} Histogram (circles)
of the tunnel-out times for spin-up electrons,
$\tau_{\uparrow,\mathrm{out}}$ (\textbf{d}), and subsequent
tunnel-in times for spin-down electrons,
$\tau_{\downarrow,\mathrm{in}}$ (\textbf{e}), as defined on the top
trace in panel \textbf{a}. In \textbf{d}, notice a systematic $\sim
10$~$\mu$s delay between the beginning of the Read phase and the
tunnel-out events, due to the response of the amplifier and filter
chain. The solid lines are exponential fits to extract the tunnel
rates. These values of $1/\Gamma_{\uparrow,\mathrm{out}}$ and
$1/\Gamma_{\downarrow,\mathrm{in}}$ were used to obtain the
simulated curves in panel \textbf{b}.}}
\end{figure}

To assess the effectiveness of the spin readout process for quantum
information purposes it is important to quantify the readout
fidelity, i.e., the probability that an electron spin state is
recognized correctly. In Fig.~4, we show the analysis of the readout
fidelity for a set of 10,000 traces. The spin state is decided based
on the peak value $I_{\rm p}$ taken by $I_{\rm SET}$(t) in the
interval $0 < t < 100$ $\mu$s. The probability distribution of the
peak currents $I_{\rm p}$ is shown in Fig.~4b. The two well-resolved
probability peaks indicate that $I_{\rm p}$ takes two preferential
values depending on the electron spin state. We have developed a
numerical model that accurately simulates the measurement process
and yields two separate histograms of peak current values for the
states $|\downarrow \rangle$ and $|\uparrow \rangle$,
$N_{\downarrow,\uparrow}(I_{\rm p})$, respectively (see
Supplementary Information). The calculated
$N_{\downarrow,\uparrow}(I_{\rm p})$ are in excellent agreement with
the measured histogram (Fig.~4b). With the knowledge of
$N_{\downarrow,\uparrow}(I_{\rm p})$, the readout
fidelities\cite{barthel09PRL} are obtained as $F_{\downarrow} = 1
\int_{I_{\rm T}}^{\infty} N_{\downarrow}(I) \mathrm{d}I$ and
$F_{\uparrow} = 1 \int_{-\infty}^{ I_{\rm T}} N_{\uparrow}(I)
\mathrm{d}I$ for the states $|\downarrow \rangle$ and $|\uparrow
\rangle$, respectively, as a function of the discrimination
threshold $I_{\rm T}$ (Fig.~4c). The integrals in
$F_{\downarrow,\uparrow}$ represent the probability that the spin
state is incorrectly assigned, either because a spin-down trace has
a noise spike $> I_{\rm T}$, or because a spin-up signal does not
reach the threshold. The visibility, defined as $V = F_{\downarrow}
+ F_{\uparrow} - 1$, reaches a maximum value $\approx 92$\% at
$I_{\rm T} = 1.1$~nA, where the readout fidelities are $
F_{\downarrow} \approx 99$\% and $ F_{\uparrow}  \approx 93$\%.

Combining spin resonance experiments\cite{morello09PRB,xiao04N} with
the ability to read out a single spin will provide a definitive
framework in which to  demonstrate and exploit coherent quantum
control of a donor electron and nuclear spin. The high-fidelity,
single-shot electron spin readout, demonstrated here for the first
time in silicon, represents the critical step to unlock the full
potential of silicon-based quantum information science and
technology.

\begin{acknowledgments}
We thank D.D. Awschalom, C. Tahan, J.J.L. Morton and G.
Prawiroatmodjo for comments and suggestions, and R.P. Starrett, D.
Barber, A. Cimmino and R. Szymanski for technical assistance. We
acknowledge support from the Australian Research Council, the
Australian Government, the U.S. National Security Agency and the
U.S. Army Research Office under Contract No. W911NF-08-1-0527.

$^*$ Electronic address: a.morello@unsw.edu.au

\dag~Present address: Walther-Meissner-Institut, Bayerische Akademie
der Wissenschaften, 85748 Garching, Germany.

\ddag~Present address: Department of Physics, University of Illinois
at Urbana-Champaign, Urbana IL 61801, USA.
\end{acknowledgments}


\end{document}